\documentclass[vecphys]{svmult}

\usepackage{makeidx}       
\usepackage{graphicx}      
                         
\usepackage{multicol}        
\usepackage[bottom]{footmisc}

\makeindex

\begin{document}

\title*{ The Shape and Topology of the Universe }

\author{Jean-Pierre Luminet}

\institute{Laboratoire Univers et Th\'eories (LUTH), Observatoire de Paris, CNRS, Universit\'e Paris Diderot, 5 place Jules Janssen, 92190 Meudon, France
\texttt{jean-pierre.luminet@obspm.fr}}

\maketitle

What is the shape of the Universe? Is it curved or flat, finite or infinite ? Is space ``wrapped around'' to create ghost images of faraway cosmic sources? We review how tessellations allow to build multiply-connected 3D Riemannian spaces useful for cosmology. We discuss more particularly the proposal of a finite, positively curved, dodecahedral space for explaining some puzzling features of the cosmic microwave background radiation, as revealed by the 2003-2006 WMAP data releases.

\section{The Hall of Mirrors}
\label{sec:1}

Imagine a room paneled with mirrors on all four vertical walls, and place ourselves somewhere within the room: a kaleidoscopic effect will be produced in the closest corner. Moreover, the repeated reflections of each pair of opposing mirrors ceaselessly reproduce the effect, creating the illusion of an infinite network extending in a plane. This paving of an infinite plane by a repeating design is called a {\it tessellation} ({\it tessella} being the name for a mosaic tile) of the Euclidean plane.

Let us now consider a room paneled with mirrors on all six surfaces (including the floor and the ceiling). If we go into the room, the interplay of multiple reflections will immediately cause us to have the impression of seeing infinitely far in every direction. Cosmic space, which is seemingly gigantic, might be lulling us with a similar illusion. Of course, it possesses neither walls nor mirrors, and the ghost images would be created not by the reflection of light from the surface of the Universe, but by a multiplication of the light ray trajectories following the folds of a wraparound universe. We could live in a physical space which is closed, small and multiply-connected, yet have the illusion that the observed space is greater, as a part of a tessellation built on repetitions of a fundamental cell. Treating such global aspects of space requires a mixture of advanced mathematics and subtle cosmological observations : Cosmic Topology~\cite{lalu}.

\begin{figure}[h!]
\centering
\includegraphics[width=11cm]{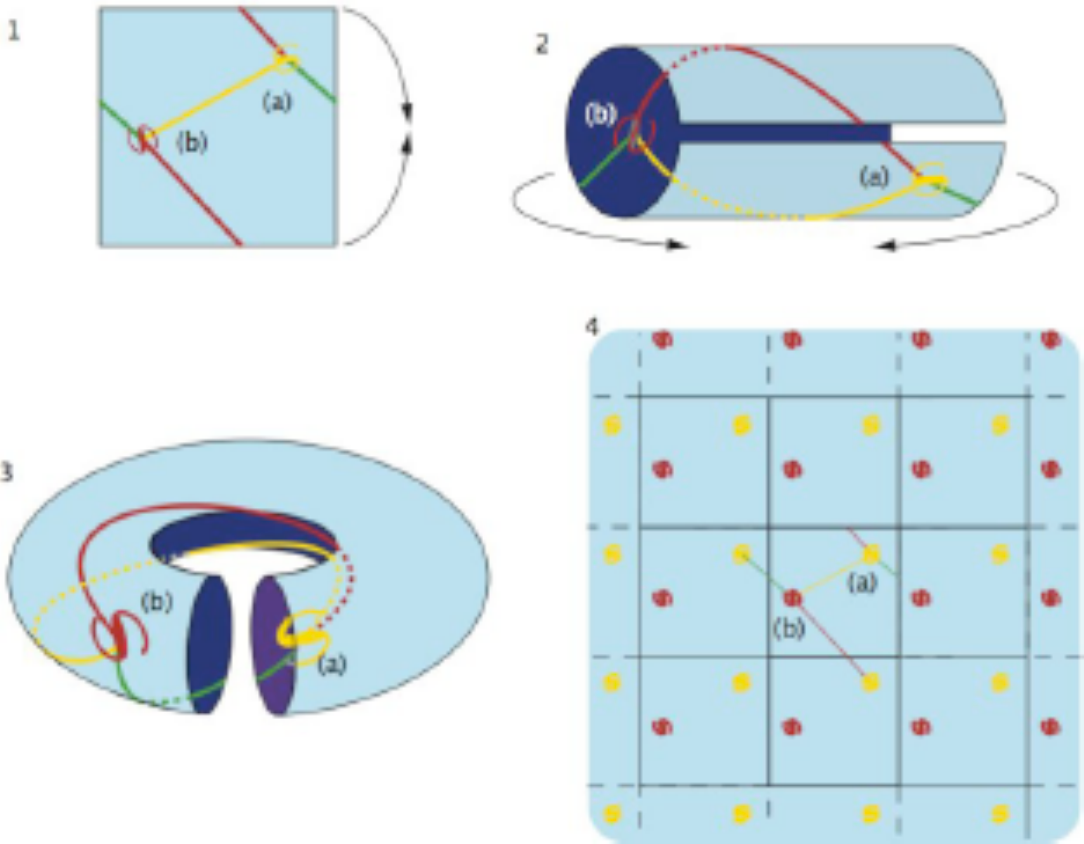}
\caption{A very simple, so-called toric universe in two dimensions, (3), shows how an observer situated in galaxy (b) can see multiple images of galaxy (a). This model of a ``wraparound'' universe is constructed by starting with a square (1), whose opposite borders have been glued together (2): everything which leaves on one side reappears immediately on the opposite side, at the corresponding point. The light from galaxy (a) reaches the galaxy (b) by several distinct trajectories, because of which the observer in galaxy (b) sees numerous images of the galaxy (a), spread in all directions across the sky. Although the space of the torus is finite, a being who lives there has the illusion of seeing a space that, if not infinite (in practice, there are horizons which limit the view), at least seems larger than it really is. This fictional space (4) looks like a network tessellated from a fundamental cell, which endlessly repeats each of the objects within the cell.}
\label{figHall}       
\end{figure}

\section{Tessellations and Topology }
\label{sec:2}

There are two complementary aspects of geometry as the science of space: the {\it metric} part deals with the properties of distance, while the {\it topological} part studies the global properties, without introducing any measurements. The topological properties are those which remain insensitive to deformations, provided that these are continuous. 

Let us take the Euclidean plane : its local geometry is determined by the metric, i.e. the way in which lengths are measured. Here, it is sufficient to apply the Pythagorean theorem for a system of two rectilinear coordinates covering the plane : $ds^2 = dx^2 + dy^2$. This is a local measurement which says nothing about the finite or infinite character of space. Now let us change the topology. To do so, we cut a strip of infinite length in one direction and finite width in the other, then we glue the two sides of the strip: we obtain a cylinder. In this operation, the metric has not changed : the Pythagorean theorem still holds for the surface of the cylinder. Nevertheless, the cylinder has a different topology : its most remarkable characteristic is the existence of an infinite number of ``straight lines'' which join two arbitrary distinct points (viewed in three dimensions, they are helices with constant spacing).

Now take a rectangle and glue its opposite edges two by two. We obtain a {\it flat torus}, a surface whose global properties are identical to those of a ring but whose curvature is everywhere zero. The metric (local geometry) of the flat torus is still given by the Pythagorean theorem, just like that of the plane and the cylinder. But the global shape is radically different, since space is now of finite extent. 

\begin{figure}[h!]
\centering
\includegraphics[width=11cm]{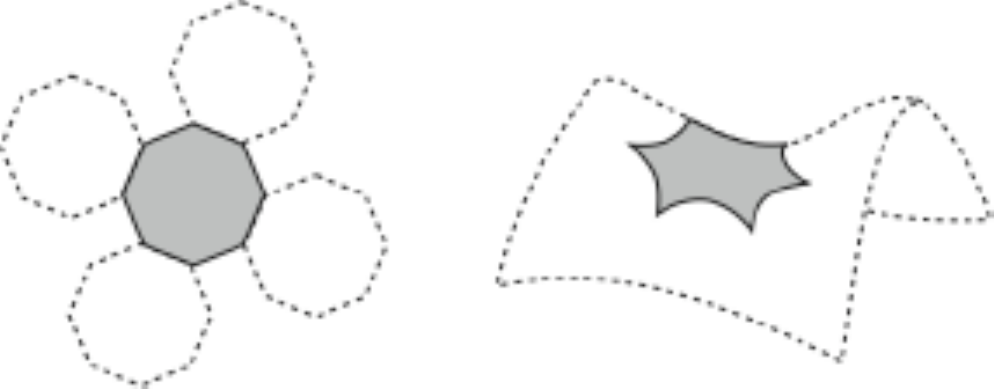}
\caption{ Paving the hyperbolic plane with octagons. It is impossible to tessellate the Euclidean plane with octagons, which implies that the double torus is not a Euclidean surface. On the other hand, the hyperbolic plane can be paved by octagons cut from the hollow of a saddle. The hyperbolic plane is thus the universal covering space for the double torus. The eight corners of the octagon must all be identified as a single point; this is the reason why one must use a negatively curved octagon with angles of $45^{\circ}$ ($8 \times 45 = 360$), in place of a flat octagon, whose angles are each $135^{\circ}$.}
\label{figHyper}       
\end{figure}

Through simple cutting and re-gluing of parts of the plane, we have thus defined two surfaces with different topologies than the plane: the cylinder and the flat torus, which however belong to the same family, the {\it locally Euclidean surfaces}. The gluing method becomes extremely fruitful when the surfaces are more complicated. Let us take two tori and glue them to form a ``double torus''. As far as its topological properties are concerned, this new surface with two holes can be represented as an eight-sided polygon (an octagon), which can be understood intuitively by the fact that each torus was represented by a quadrilateral. But this surface is not capable of tessellating the Euclidean plane, for an obvious reason: if one tries to add a flat octagon to each of its edges, the eight octagons will overlap each other. One must curve in the sides and narrow the angles, in other words pass to a hyperbolic space: only there does one succeed in fitting eight octagons around the central octagon, and starting from each of the new octagons one can construct eight others, {\it ad infinitum}. By this process one tessellates an infinite space : the Lobachevsky hyperbolic plane (Fig.~\ref{figHyper}).

\begin{figure}[h!]
\centering
\mbox{\hskip 0.0truecm\includegraphics[width=9cm]{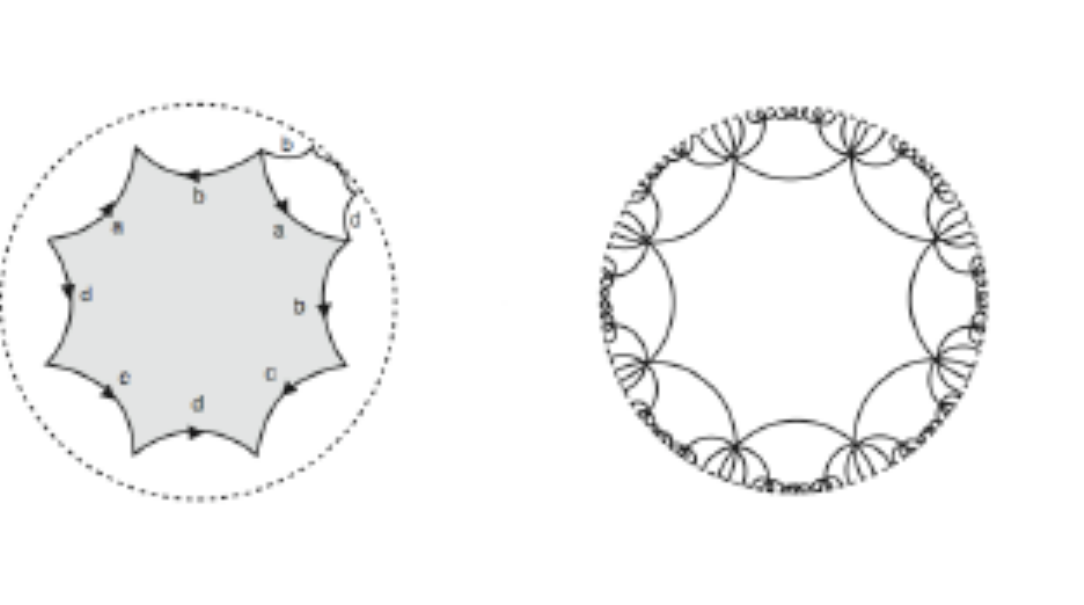}}
\mbox{\hskip 0.5truecm\includegraphics[width=7cm]{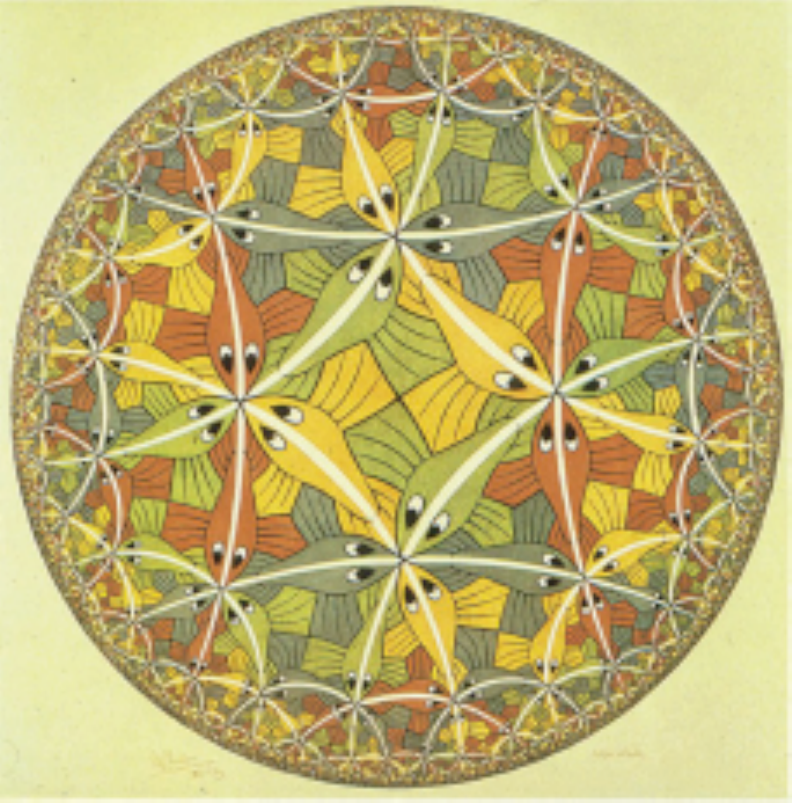}}
\caption{ \textit{Upper} : Poincar\'e's representation of the hyperbolic plane. By acting with the holonomies on each point of the fundamental octagon, and repeating the process again and again, one creates a tessellation of the hyperbolic plane by regular and identical octagons. Poincar\'e demonstrated that the hyperbolic plane, normally infinite, could be represented entirely within the interior of a disk, whose edge represents infinity. Poincar\'e's model deforms distances and shapes, which explains why the octagons seem irregular and increasingly tiny as we approach the boundary of the disk. All of the lines in the figure represent straight lines of the hyperbolic plane, and meet the boundary at a right angle. \textit{Lower} : In this 1959 woodcutting entitled Circle Limit III, Escher has used the representation given by Poincar\'e to tessellate the hyperbolic plane using fish.}
\label{figHyppoincare}       
\end{figure}

A fascinating representation of a hyperbolic tessellation was given by Poincar\'e. A conformal change of coordinates allows us to bring infinity to a finite distance, with the result that the entire Lobachevsky space is contained in the interior of the unit disk. The famous Dutch graphic artist Maurits Cornelis Escher created a series of prints entitled {\it Circle Limit}, in which he used Poincar\'e's representation (see Fig.~\ref{figHyppoincare}).

More generally, a torus with $n$ holes, ${\bf T}_n$, can be constructed as the connected sum of $n$ simple tori. It is topologically equivalent to a $4n$-gon where all the vertices are identical with each other and the sides are suitably identified by pairs. The $n$-torus ($n \ge 2$) is a compact surface of negative curvature. This type of surface is most commonly seen at bakeries, in the form of pretzels. They all have the same local geometry, of hyperbolic type ; however, they do not have the same topology, which depends on the number of holes.

Thus it is possible to represent any surface whatsoever with a polygon whose sides one identifies, two by two. The polygon is called a {\it fundamental domain} (hereafter FD). The FD distinctly characterizes a certain aspect of the topology. But this is not enough; we must also specify the geometric transformations which identify the points. Indeed, starting from a square, one could identify the points diametrically opposite with respect to the center of symmetry of the square, and the surface obtained will no longer be a flat torus; it will no longer even be Euclidean, but spherical, a surface called the projective plane. The mathematical transformations used to identify points form a group of symmetries, called the {\it holonomy group}. If the holonomy group is trivial, the space is {\it simply connected}. If not, it is {\it multiply connected}.

The holonomy group is discrete, i.e. there is a non zero shortest distance between any two homologous points, and the generators of the group (except the identity) have no fixed point. This last property is very restrictive (it excludes for instance the rotations) and allows the classification of all possible holonomy groups. Due to the fact that the holonomy group is discrete, the FD is always convex and has a finite number of faces. In two dimensions, it is a surface whose boundary is constituted by lines, thus a polygon. In three dimensions, it is a volume bounded by faces, thus a polyhedron.

Starting from the fundamental domain and acting with the transformations of the holonomy group on each point, one creates a number of replicas of the FD ; we produce a tessellation of a larger space, called the {\it universal covering space} (hereafter UC) $\bf{M}*$. By construction, $\bf{M}*$ is locally indistinguishable from $\bf{M}$. But its topological properties can be quite different. The UC is necessarily simply connected. When $\bf{M}$ is multiply connected, each point of $\bf{M}$ generates replicas of points in $\bf{M}*$. The universal covering space can be thought of as an unwrapping of the original manifold. For instance, the UC of the flat torus is the Euclidean plane ${\bf E}^2$, which indeed reflects the fact that the flat torus is a locally Euclidean surface. 

The shape of a homogeneous space is entirely specified if one is given a fundamental domain; a particular group of symmetries, the holonomies (fixed point-free discrete subgroup of isometries), which identify the edges of the domain two by two; and a universal covering space that is tessellated by fundamental domains. Classifying the possible shapes thus reduces, in part, to classifying symmetries.

\section{Species of Spaces}
\label{sec:3}

Cosmological solutions of general relativity focus mainly on locally homogeneous and isotropic spaces, namely those admitting one of the three geometries of constant curvature. Any compact 3-manifold $\bf{M}$ with constant curvature $k$ can thus be expressed as the quotient $\bf{M} = \bf{M}*/\bf{G}$  where the universal covering space $\bf{M}*$ is either :
\begin{itemize}
\item the Euclidean space $\bf{E}^3$ if $k = 0$
\item  the hypersphere $\bf{S}^3$ if $k > 0$
\item the hyperbolic 3-space $\bf{H}^3$ if $k < 0$
\end{itemize}
and the holonomy group $\bf{G}$ is a subgroup of isometries of $\bf{M}*$ acting freely and discontinuously.

Given the recent observational constraints on the curvature of cosmic space (see below), in the remaining of this short review we focus our attention to Euclidean and spherical spaces only. 

\begin{figure}[h!]
\centering
\mbox{\hskip 0.0truecm\includegraphics[width=5.5cm]{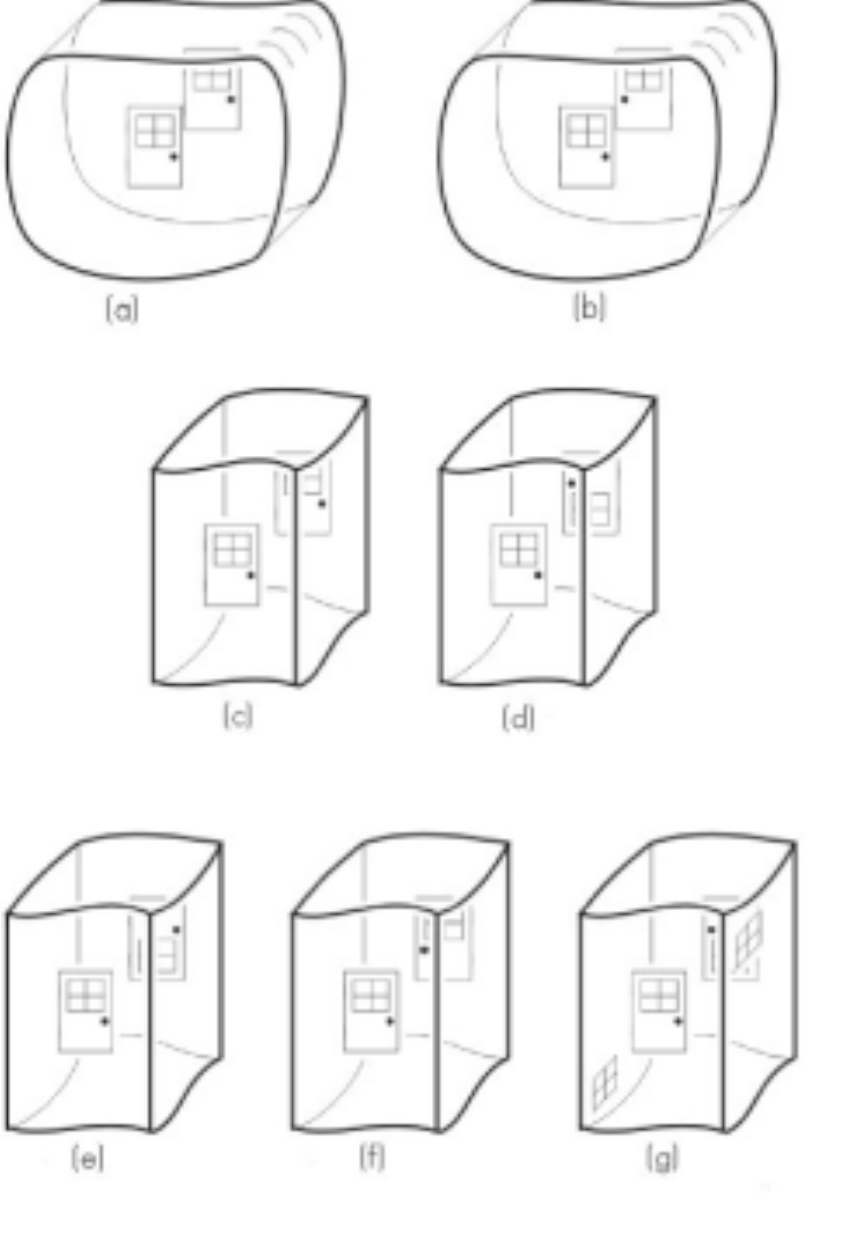}}
\mbox{\hskip 0.5truecm\includegraphics[width=5.5cm]{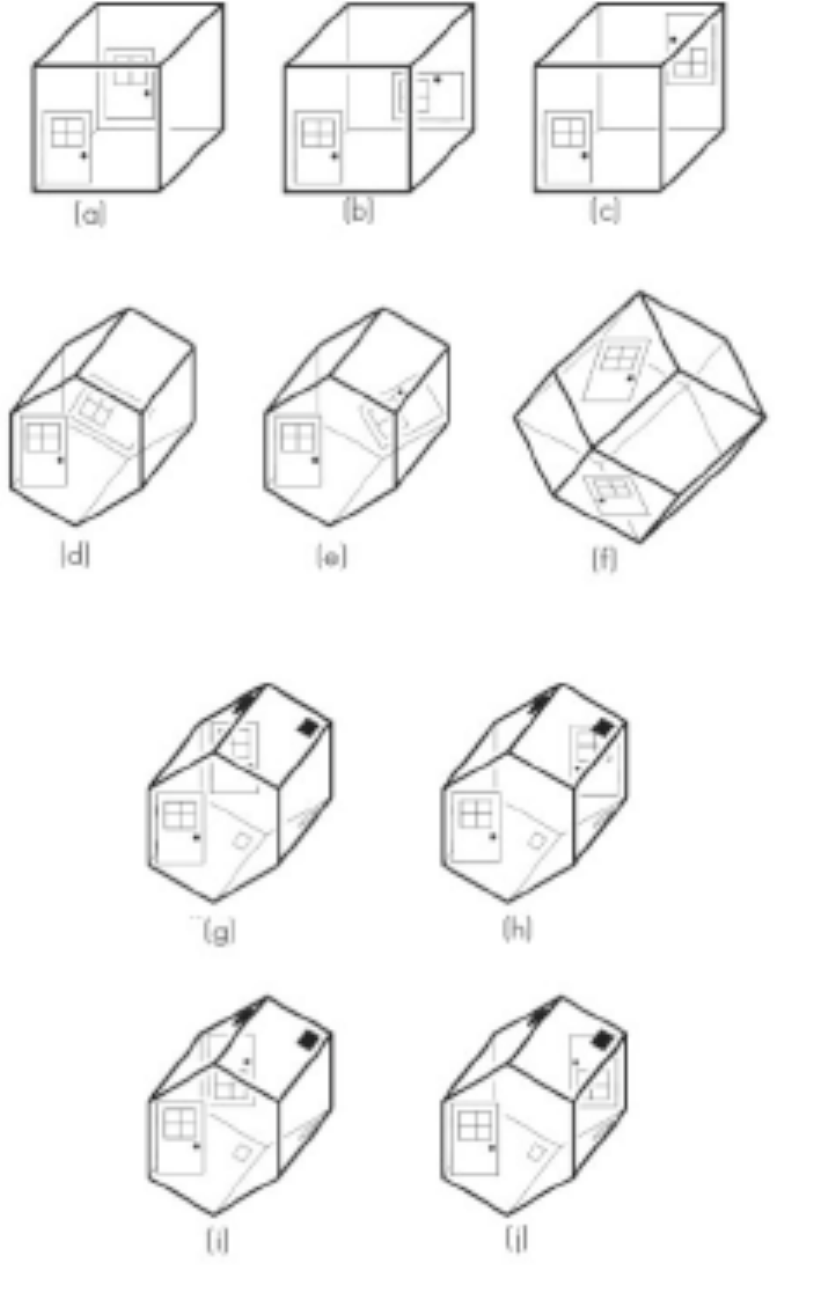}}
\caption{The Seventeen Multiply-Connected Euclidean Spaces. The orientation of the doors indicates how the corresponding walls must be glued together (courtesy Adam Weeks Marano).}
\label{fig17flat}      
\end{figure}

The multiply connected Euclidean spaces are characterized by their fundamental polyhedra and their holonomy groups. The fundamental polyhedra are either a finite or infinite parallelepiped, or a prism with a hexagonal base, corresponding to the two ways of tessellating Euclidean space. The various combinations generate seventeen multiply connected Euclidean spaces, as shown in Fig.~\ref{fig17flat} (for an exhaustive study, see \cite{Ria04a}).
Seven of these spaces (called slabs and chimneys) are of infinite volume. The ten other are of finite volume, six of them being orientable hypertori. The latter present a particular interest for cosmology, since they could perfectly model the spatial part of the so-called ``flat'' universe models.

\begin{figure}[h!]
\centering
\includegraphics[width=11cm]{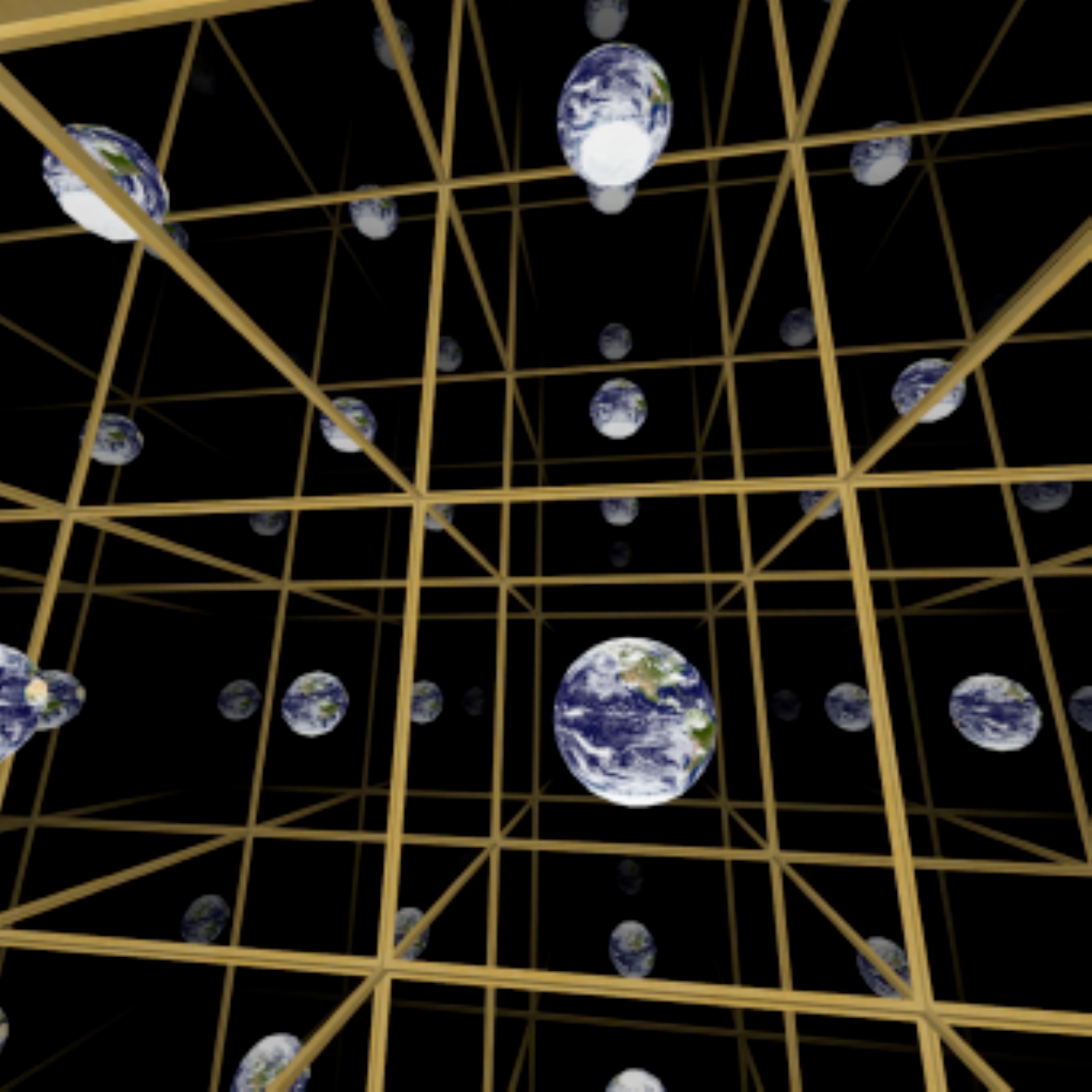}
\caption{ By analogy with the two-dimensional case, the three-dimensional hypertorus $\bf{T}^3$ is obtained by identifying the opposite faces of a parallelepiped. The resulting volume is finite. Let us imagine a light source at our position, immersed in such a structure. Light emitted backwards crosses the face of the parallelepiped behind us and reappears on the opposite face in front of us; therefore, looking forward we can see our back. Similarly, we see in our right our left profile, or upwards the bottom of our feet. In fact, for light emitted isotropically, and for an arbitrarily large time to wait, we could observe ghost images of any object (here the Earth) viewed arbitrarily close to any angle. The resulting visual effect would be comparable (although not identical) to what could be seen from inside a parallelepiped of which the internal faces are covered with mirrors. Thus one would have the visual impression of infinite space, although the real space is closed (courtesy Jeff Weeks).}
\label{fighypertorus}      
\end{figure}

For spherical spaces, the simply-connected hypersphere $\bf{S}^3$ can be viewed as composed of two spherical balls embedded in Euclidean space, glued along their boundaries in such a way that each point on the boundary of one ball is the same as the corresponding point on the other ball. The full isometry group of $\bf{S}^3$ is $SO(4)$. The holonomies that preserve the metric of the hypersphere, i.e. the admissible subgroups $\bf{G}$ of $SO(4)$ without fixed point, acting freely and discontinuously on $\bf{S}^3$, belong to three categories :
\begin{itemize}
\item the cyclic groups of order $p$, $\bf{Z}_p$ ($p \ge 2$), made up of rotations by an angle $2\pi/p$ around a given axis, where $p$ is an arbitrary integer ;
\item the dihedral groups of order $2m$, $\bf{D}_m$ ($m > 2$), which are the symmetry groups of a regular plane polygons of $m$ sides;
\item the binary polyhedral groups, which preserve the shapes of the regular polyhedra. The group $\bf{T}*$ preserves the tetrahedron (4 vertices, 6 edges, 4 faces), of order 24 ; the group $\bf{O}*$ preserves the octahedron (6 vertices, 12 edges, 8 faces), of order 48 ; the group $\bf{I}*$ preserves the icosahedron (12 vertices, 30 edges, 20 faces), of order 120. There are only three distinct polyhedral groups for the five polyhedra, because the cube and the octahedron on the one hand, the icosahedron and the dodecahedron on the other hand are duals, so that  their symmetry groups are the same.
\end{itemize}

If one identifies the points of the hypersphere by holonomies belonging to one of these groups, the resulting space is spherical and multiply connected. For an exhaustive classification, see \cite{Gaus01}. There is a countable infinity of these, because of the integers $p$ and $m$ which parametrize the cyclic and dihedral groups. 

Since the universal covering $\bf{S}^3$ is compact, all the multiply connected spherical spaces are also compact. As the volume of $\bf{S}^3$ is $2 \pi^2R^3$, the volume of $\bf{M} = \bf{S}^3/\bf{G}$ is simply $vol({\bf M}) = 2 \pi^2R^3/G$ where $G$ is the order of the group $\bf{G}$. For topologically complicated spherical 3-manifolds, $G$ becomes large and $vol(\bf{M})$ is small. There is no lower bound since $\bf{G}$ can have an arbitrarily large number of elements (for lens and prism spaces, the larger $p$ and $m$ are, the smaller the volume of the corresponding spaces). Hence $0 < vol({\bf M}) \le 2 \pi^2R^3$. In contrast, the diameter, i.e., the maximum distance between two points in the space, is bounded below by $\simeq 0.326 R$, corresponding to the dodecahedral space.

\begin{figure}[h!]
\centering
\mbox{\hskip 0.0truecm\includegraphics[width=5.5cm]{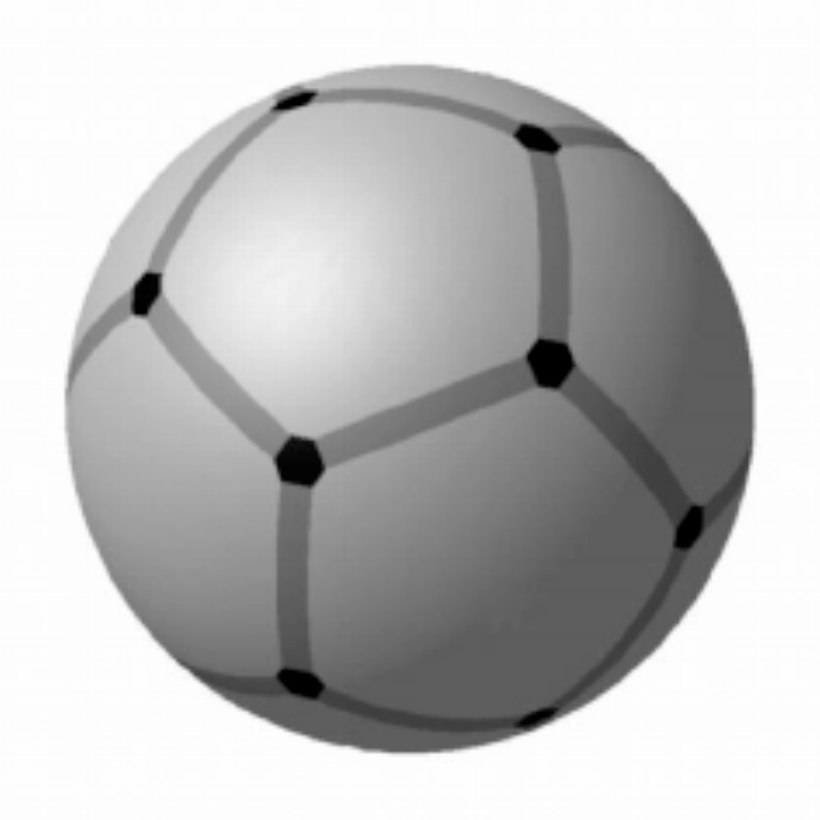}}
\mbox{\hskip 0.5truecm\includegraphics[width=5.5cm]{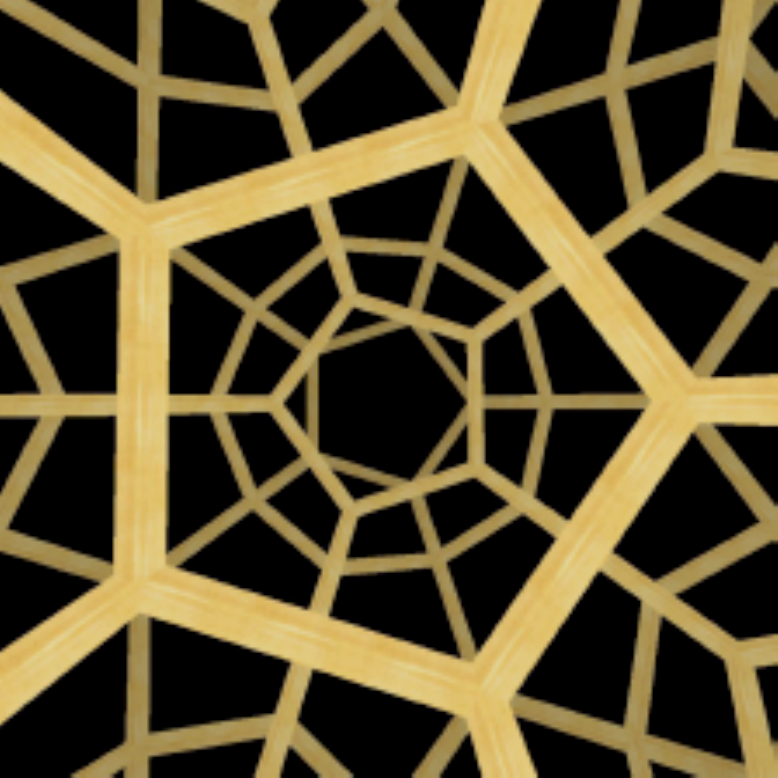}}
\caption {\textit{Left} : Poincar\'e Dodecahedral Space can be described as the interior of a spherical dodecahedron such that when one goes out from a pentagonal face, one comes back immediately inside the space from the opposite face, after a $36^{\circ}$ rotation. Such a space is finite, although without edges or boundaries, so that one can indefinitely travel within it. \textit{Right} : View from inside PDS perpendicularly to one pentagonal face. In such a direction, ten dodecahedra tile together with a $1/10^{th}$ turn to tessellate the universal covering space $\bf{S^3}$. Since the dodecahedron has 12 faces, 120 dodecahedra are necessary to tessellate the full hypersphere. Thus, an observer has the illusion to live in a space 120 times vaster, made of tiled dodecahedra which duplicate like in a mirror hall (courtesy Jeff Weeks).}
\label{figsoccer}      
\end{figure}

\begin{figure}[h!]
\centering
\includegraphics[width=11cm]{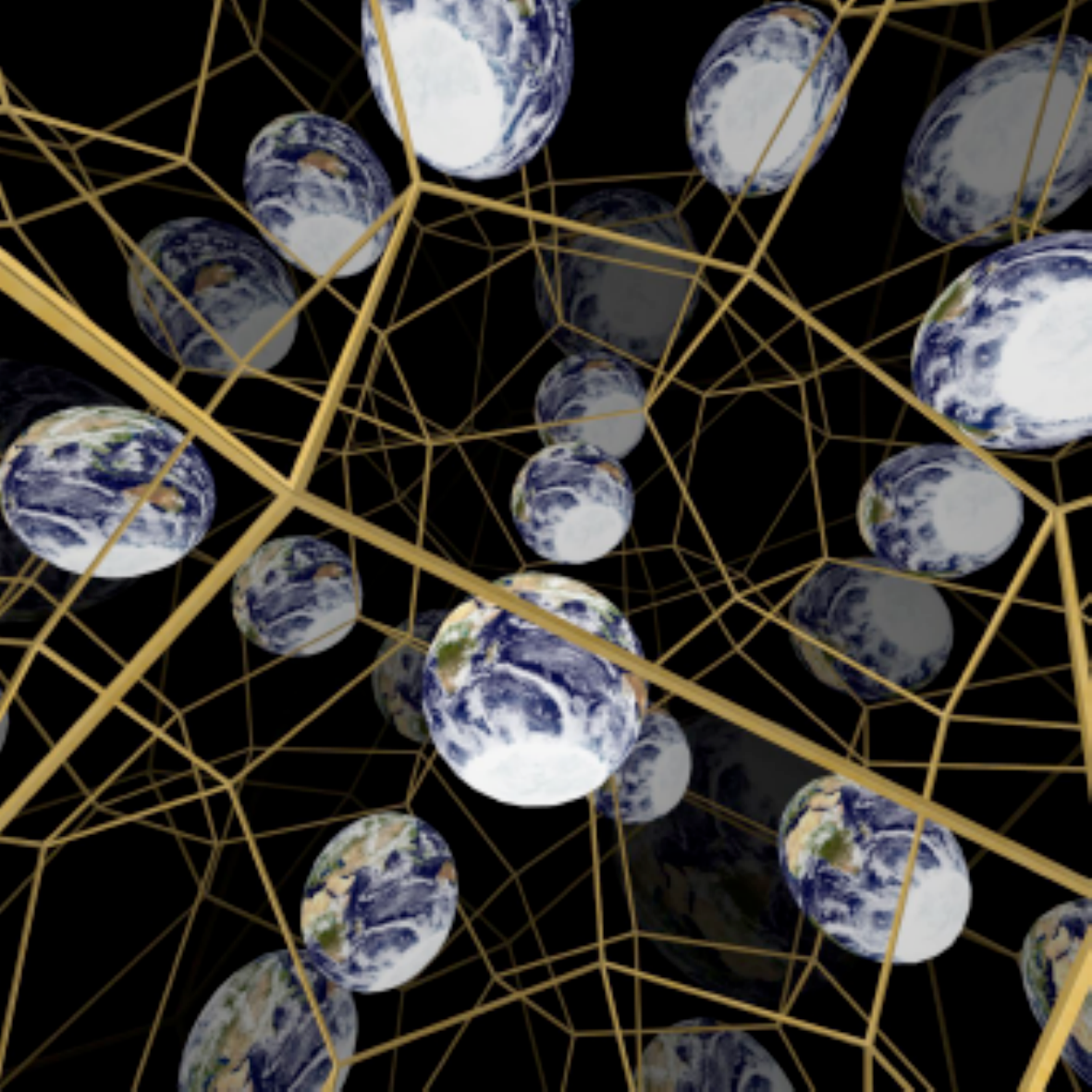}
\caption{ View from inside PDS calculated by the CurvedSpaces program (courtesy Jeff Weeks).}
\label{figPDSinside2}     
\end{figure}

Let us now concentrate on the properties of the {\it Poincar\'e Dodecahedral Space} $\bf{S}^3/\bf{I}*$ (hereafter PDS), obtained by identifying the opposite pentagonal faces of a regular spherical dodecahedron after rotating by $36^{\circ}$ in the clockwise direction around the axis orthogonal to the face (Fig.~\ref{figsoccer}). This configuration involves 120 successive operations and gives some idea of the extreme complication of such multiply connected topologies. Its volume is 120 times smaller than that of the hypersphere with the same radius of curvature, and it is of particular interest for cosmology, giving rise to fascinating topological mirages (Fig.~\ref{figPDSinside2}).

\begin{figure}[h!]
\centering
\includegraphics[height=3.5cm]{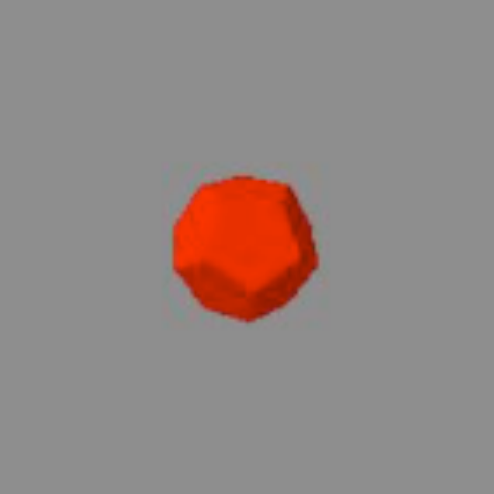}
\includegraphics[height=3.5cm]{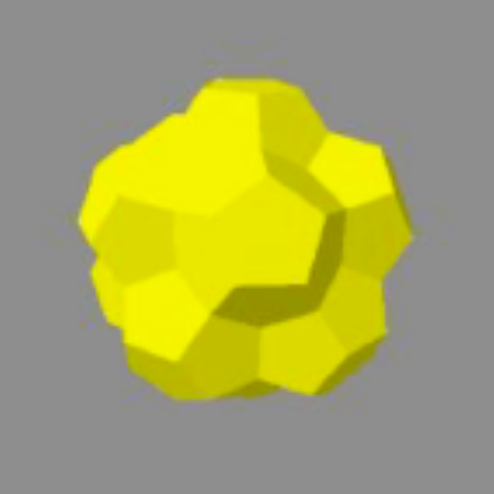}
\includegraphics[height=3.5cm]{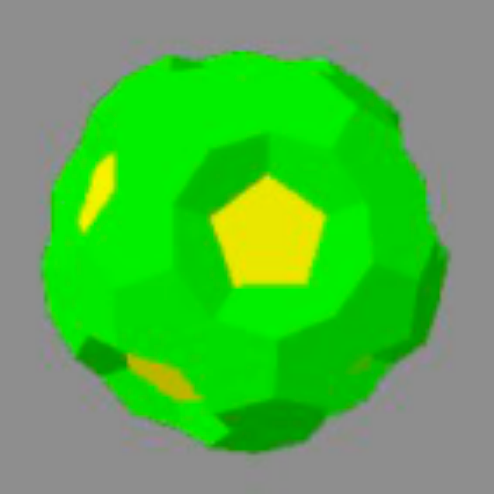}
\includegraphics[height=3.5cm]{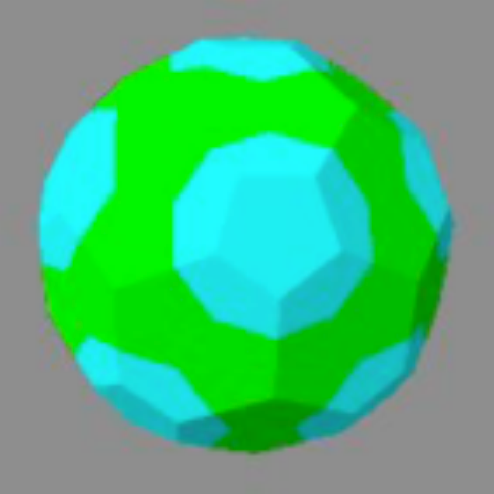}
\includegraphics[height=3.5cm]{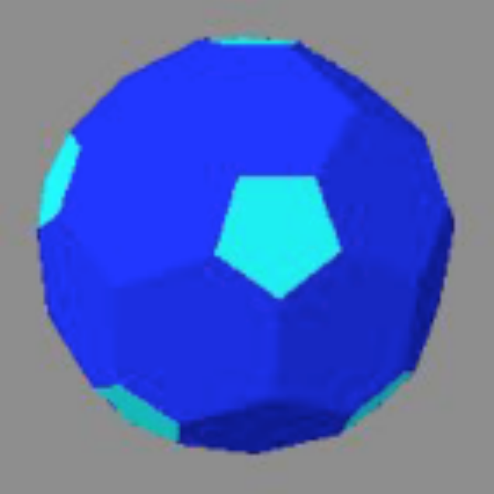}
\caption{ The first 5 layers of PDS (courtesy Jeff Weeks).}
\label{figPDSlayers}      
\end{figure}

The tessellation of $\bf{S}^3$ by 120 copies of the PDS is not obvious to visualize, see Fig.~\ref{figPDSlayers}. It involves 9 successive layers. Start with the original cell (layer 1). It has 12 pentagonal faces, on each of them one builds a spherical dodecahedron, thus we get 12 dodecahedra on layer 2.  Then we go further on and we get 20 dodecahedra in layer 3, 12 in layer 4, 30 in layer 5, 12 in layer 6, 20 in layer 7, 12 in layer 8 and 1 in layer 9 (it can be checked that the sum is 120). Layers 6 through 9 are of course symmetrical to layers 4 through 1. Note that the cells in layer 5 sit ``vertically'' with respect to the equatorial hyperplane (i.e. they are orthogonal to the equatorial hyperplane), which is why they appear flat in the image (each dark blue hexagon is the 2D shadow of a 3D cell when it is projected from 4D space to 3D space).

\section{Topology and Cosmology}
\label{sec:4}

It is presently believed that our Universe is correctly described at large scale by a Friedmann-Lema\^{\i}tre (hereafter FL) model. The FL models are homogeneous and isotropic solutions of Einstein's equations, of which the spatial sections have constant curvature. The FL models fall into 3 general classes, according to the sign of their spatial curvature $k= -1, 0, +1$. The spacetime manifold is described by the metric $ds^2 = c^2dt^2 - R^2(t)d\sigma^2$, where $d\sigma^2 = d\chi^2 + {S_k}^2(\chi) (d\theta^2+\sin^2{\theta}d\varphi^2)$ is the metric of a 3-dimensional homogeneous manifold, flat [$k=0$] or with curvature [$k \pm 1$]. The function $S_k(\chi)$ is defined as $\sinh(\chi)$ if $k= -1$, $\chi$ if $k=0$, $\sin(\chi)$ if $k=1$; $R(t)$ is the scale factor, chosen equal to the spatial curvature radius for non flat models.

In most studies, the spatial topology is assumed to be that of the corresponding simply connected space: the hypersphere, Euclidean space or the 3D-hyperboloid, the first being finite and the other two infinite. However, there is no particular reason for space to have a simply connected topology. In any case, general relativity says nothing on this subject; it is only the strict application of the cosmological principle, added to the theory, which encourages a generalization of locally observed properties to the totality of the Universe. However, to the metric element given above there are several, if not an infinite number, of possible topologies, and thus of possible models for the physical Universe. For example, the hypertorus and the familiar Euclidean space are locally identical, and relativistic cosmological models describe them with the same FL equations, even though the former is finite and the latter infinite. Only the boundary conditions on the spatial coordinates are changed. Thus the multiply connected cosmological models share exactly the same kinematics and dynamics as the corresponding simply connected ones (for instance, the time evolutions of the scale factor $R(t)$ are identical). 

In FL models, the curvature of physical space depends on the way the total energy density of the Universe may counterbalance the kinetic energy of the expanding space. The normalized density parameter $\Omega_0$, defined as the ratio of the actual energy density to the critical value that an Euclidean space would require, characterizes the present-day contents (matter, radiation and all forms of energy) of the Universe. If $\Omega_0$ is greater than 1, then space curvature is positive and geometry is spherical; if $\Omega_0$ is smaller than 1 the curvature is negative and geometry is hyperbolic; eventually $\Omega_0$ is strictly equal to 1 and space is locally Euclidean. 

The next question about the shape of the Universe is to know whether space is finite or infinite - equivalent to know whether space contains a finite or an infinite amount of matter-energy, since the usual assumption of homogeneity implies a uniform distribution of matter and energy through space. From a purely geometrical point of view, all positively curved spaces are finite whatever their topology, but the converse is not true : flat or negatively curved  spaces can have finite or infinite volumes, depending on their degree of connectedness (see e.g. \cite{Ell71},\cite{lalu}).

From an astronomical point of view, it is necessary to distinguish between the ``observable universe'', which is the interior of a sphere centered on the observer and whose radius is that of the cosmological horizon (roughly the radius of the last scattering surface), and the physical space. There are only three logical possiblities. First, the physical space is infinite - like for instance the simply connected Euclidean space. In this case, the observable universe is an infinitesimal patch of the full universe and, although it has long been the preferred model of many cosmologists, this is not a testable hypothesis. Second, physical space is finite (e.g. an hypersphere or a closed multiply connected space), but greater than the observable space. In that case, one easily figures out that if physical space is much greater that the observable one, no signature of its finitude will show in the observable data. But if space is not too large, or if space is not globally homogeneous (as is permitted in many space models with multiply connected topology) and if the observer occupies a special position, some imprints of the space finitude could be observable. Third, physical space is smaller than the observable universe. Such an apparently odd possibility is due to the fact that space can be multiply connected and have a small volume. There is a lot of possibilites, whatever the curvature of space. Small universe models may generate multiple images of light sources, in such a way that the hypothesis can be tested by astronomical observations. The smaller the fundamental domain, the easier it is to observe the multiple topological imaging. How do the present observational data constrain the possible multi-connectedness of the universe and, more generally, what kinds of tests are conceivable ? (see \cite{Lum07} for a non-technical book about all the aspects of topology and its applications to cosmology). 

If the Universe was finite and small enough, we should be able to see ``all around'' it, because the photons might have crossed it once or more times. In such a case, any observer might identify multiple images of a same light source, although distributed in different directions of the sky and at various redshifts, or to detect specific statistical properties in the apparent distribution of faraway sources such as galaxy clusters. To do this, methods of {\it cosmic crystallography} have been devised (see e.g.\cite{lelalu96}, \cite{ull99}). The main limitation of cosmic crystallography is that the presently available catalogs of observed sources at high redshift are not complete enough to perform convincing tests. 

Fortunately, the topology of a small Universe may also be detected through its effects on such a Rosetta stone of cosmology as is the cosmic microwave background (hereafter CMB) fossil radiation (for a review, \cite{Levin02}). If you sprinkle fine sand uniformly over a drumhead and then make it vibrate, the grains of sand will collect in characteristic spots and figures, called Chladni patterns. These patterns reveal much information about the size and the shape of the drum and the elasticity of its membrane. In particular, the distribution of spots depends not only on the way the drum vibrated initially but also on the global shape of the drum, because the waves will be reflected differently according to whether the edge of the drumhead is a circle, an ellipse, a square, or some other shape. In cosmology, the early Universe was crossed by real acoustic waves generated soon after the big bang. Such vibrations left their imprints 380 000 years later as tiny density fluctuations in the primordial plasma. Hot and cold spots in the present-day 2.7 K CMB radiation reveal those density fluctuations. Thus the CMB temperature fluctuations look like Chladni patterns resulting from a complicated three-dimensional drumhead that vibrated for 380 000 years. They yield a wealth of information about the physical conditions that prevailed in the early Universe, as well as present geometrical properties like space curvature and topology. More precisely, density fluctuations may be expressed as combinations of the vibrational modes of space, just as the vibration of a drumhead may be expressed as a combination of the drumhead's harmonics. The shape of space can be heard in a unique way. Lehoucq et al.~\cite{Leh02} calculated the harmonics (the so-called ``eigenmodes of the Laplace operator'') for most of the spherical topologies, and Riazuelo et al.~\cite{Ria04a} did the same for all 18 Euclidean spaces. Then, starting from a set of initial conditions fixing how the universe originally vibrated (the so-called Harrison-Zeldovich spectrum), it is possible to evolve the harmonics forward in time to simulate realistic CMB maps for a number of flat and spherical topologies~\cite{Ria04b}. 

The ``concordance model'' of cosmology describes the Universe as a flat infinite space in eternal expansion, accelerated under the effect of a repulsive dark energy. The data collected by the NASA satellite WMAP~\cite{WMAP1} have produced a high resolution map of the CMB which showed the seeds of galaxies and galaxy clusters and allowed to check the validity of the dynamic part of the expansion model. However, combined with other astronomical data~\cite{Ton03}, they suggest a value of the density parameter $\Omega_0 = 1.02 \pm 0.02$ at the $1\sigma$ level. The result is marginally compatible with strictly flat space sections. Improved measurements could indeed lower the value of $\Omega_0$ closer to the critical value 1, or even below to the hyperbolic case. Presently however, taken at their face value, WMAP data favor a positively curved space, necessarily of finite volume since all spherical spaceforms possess this property. 

CMB temperature anisotropies essentially result from density fluctuations of the primordial Universe : a photon coming from a denser region will loose a fraction of its energy to compete against gravity, and will reach us cooler. On the contrary, photons emitted from less dense regions will be received hotter. The density fluctuations result from the superposition of acoustic waves which propagated in the primordial plasma. They can be decomposed into a sum of spherical harmonics, much like the sound produced by a music instrument may be decomposed into ordinary harmonics. The ``fundamental'' fixes the height of the note, whereas the relative amplitudes of each harmonics determine the tone quality. Concerning the relic radiation, the relative amplitudes of each spherical harmonics determine the {\it power spectrum}, which is a signature of the space geometry and of the physical conditions which prevailed at the time of CMB emission.

The power spectrum depicts the minute temperature differences on the last scattering surface, depending on the angle of view. It exhibits a set of peaks when anisotropy is measured on small and mean scales (i.e. concerning regions of the sky of relatively modest size). These peaks are remarkably consistent with the infinite flat space hypothesis. At large angular scales, the first observable harmonics is the {\it quadrupole}, whose wavenumber is $\ell = 2$ (i.e. concerning CMB spots typically separated by $90^{\circ}$). The concordance model predicts that the power spectrum should follow the so-called ``Sachs-Wolfe plateau''. However, WMAP measurements fall well below the plateau : the measured value of the quadrupole is 7 times weaker than expected in a flat infinite Universe. The probability that such a discrepancy occurs by chance has been estimated to $0.2 \%$ only. The octopole (whose wavenumber is $\ell = 3$) is also weaker, but still compatible with the error bar, which is larger in this range of wavenumbers due to cosmic variance. For larger wavenumbers up to $\ell = 900$ (which correspond to temperature fluctuations at small angular scales), observations are remarkably consistent with the concordance model.

The unusually low quadrupole value means that long wavelengths are missing. Some cosmologists have proposed to explain the anomaly by still unknown physical laws of the early universe~\cite{Tsu03}. A more natural explanation may be because space is not big enough to sustain long wavelengths. Such a situation may be compared to a vibrating string fixed at its two extremities, for which the maximum wavelength of an oscillation is twice the string length. On the contrary, in an infinite flat space, all the wavelengths are allowed, and fluctuations must be present at all scales. Thus this geometrical explanation relies on a model of finite space whose size smaller than the observable universe constrains the observable wavelengths below a maximum value.

Weeks et al.~\cite{Wee04} showed that some finite multiconnected topologies do lower the large-scale fluctuations whereas others may elevate them. In fact, the long wavelengths modes tend to be relatively lowered only in a special family of closed multiconnected spaces called ``well-proportioned''. Generally, among spaces whose characteristic lengths are comparable with the radius of the last scattering surface $R_{lss}$ (a necessary condition for the topology to have an observable influence on the power spectrum), spaces with all dimensions of similar magnitude lower the quadrupole more heavily than the rest of the power spectrum. As soon as one of the characteristic lengths becomes significantly smaller or greater than the other two, the quadrupole is boosted in a way not compatible with WMAP data. In the case of flat tori, a cubic torus lowers the quadrupole whereas an oblate or a prolate torus increase the quadrupole; for spherical spaces, polyhedric spaces suppress the quadrupole whereas high order lens spaces (strongly anisotropic) boost the quadrupole. Thus, well-proportioned spaces match the WMAP data much better than the infinite flat space model.

\section{The ``football'' universe}
\label{sec:5}

Among the family of well-proportioned spaces, the best fit to the observed power spectrum is the already mentioned {\it Poincar\'e Dodecahedral Space} ~\cite{Lum03}. Recall that this space is positively curved, and is a multiply connected variant of the simply connected hypersphere $\bf{S}^3$, with a volume 120 times smaller for the same curvature radius. 

The associated power spectrum, namely the repartition of fluctuations as a function of their wavelengths corresponding to PDS, strongly depends on the value of the mass-energy density parameter. Luminet et al.~\cite{Lum03} computed the CMB multipoles for $\ell = 2, 3, 4$ and fitted the overall normalization factor to match the WMAP data at $\ell = 4$, and then examined their prediction for the quadrupole and the octopole as a function of $\Omega_0$. There is a small interval of values within which the spectral fit is excellent, and in agreement with the value of the total density parameter deduced from WMAP data ($1.02 \pm 0.02$). The best fit was obtained for $\Omega_0 = 1.016$ (with the matter component $\Omega_m = 0.28$). 

\begin{figure}[h!]
\centering
\includegraphics[width=11cm]{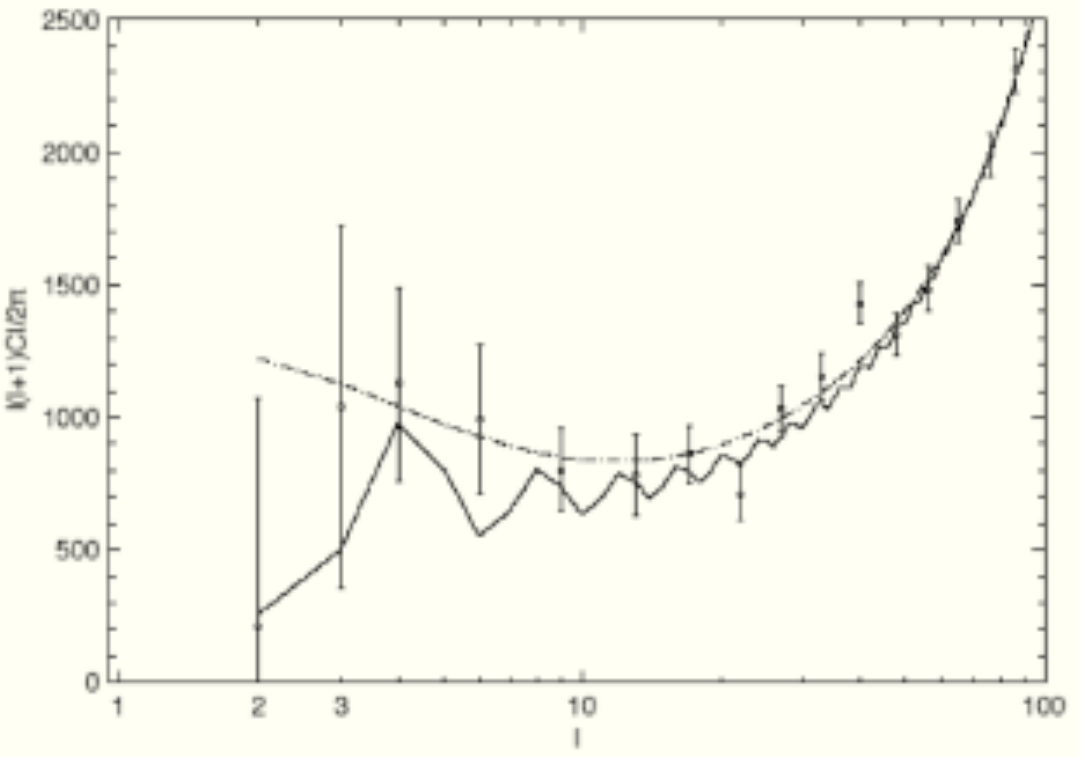}
\caption{Comparative power spectra as a function of the multipoles $\ell$ for WMAP3 (errorbars), the concordance model (dotdashed curve) and PDS (solid curve) for $\Omega_0 = 1.018$, $\Omega_{m} =0.27$ and $h = 0.70$ Here we calculate the modes up to $k = 3000$ using the conjecture of \cite{ALS_conj} proved by \cite{Gun}.}
\label{PDS07}
\end{figure}

Since then, the properties of PDS have been investigated in more details by various authors. ~\cite{SCMLR} found an analytical expression of the eigenmodes of PDS, whereas~\cite{ALS_conj} and~\cite{Gun} computed numerically the power spectrum up to the $\ell = 15$ mode (corresponding to the calculation of 10,521 eigenmodes) and showed that the fit with WMAP was obtained for $1.016 < \Omega_0 < 1.020$. More recently, Caillerie et al.~\cite{Cai07} computed the power spectrum of PDS until $\ell = 35$ (involving the calculation of $1.7 \times 10^9$ eigenmodes) and confirmed the fit (Fig.~\ref{PDS07}).

The result is quite remarkable because the Poincar\'e space has no degree of freedom. By contrast, a 3-dimensional torus, constructed by gluing together the opposite faces of a cube and which constitutes a possible topology for a finite Euclidean space, may be deformed into any parallelepiped : therefore its geometrical construction depends on 6 degrees of freedom. 

The values of the matter density $\Omega_m$, of the dark energy density $\Omega_{\Lambda}$ and of the expansion rate $H_0$ fix the radius of the last scattering surface $R_{lss}$ as well as the curvature radius of space $R_c$, thus dictate the possibility to detect the topology or not. For $\Omega_m = 0.28$, $\Omega_0 = 1.016$ and $H_0 = 62 km/s/Mpc$, $R_{lss} = 53Gpc$ and $R_c = 2.63R_{lss}$. It is to be noticed that the curvature radius $R_c$ is the same for the simply-connected universal covering space $\bf{S}^3$ and for the multiply connected PDS. Incidently, the numbers above show that, contrary to a current opinion, a cosmological model with $\Omega_0 \simeq 1.02$ is {\it far from being ``flat''} (i.e. with $R_c = \infty$) ! For the same curvature radius than the simply-connected hypersphere $\bf{S}^3$, the smallest dimension of the fundamental dodecahedron is only 43 Gpc, and its volume about $80 \%$ the volume of the observable universe (namely the volume of the last scattering surface). This implies that some points of the last scattering surface will have several copies. Such a lens effect is purely attributable to topology and can be precisely calculated in the framework of the PDS model. It provides a definite signature of PDS topology, whereas the shape of the power spectrum gives only a hint for a small, well-proportioned universe model.
 
To be confirmed, the PDS model (sometimes popularized as the ``football'' universe model) must satisfy two experimental tests :
\begin{itemize}
\item New data from the future European satellite Planck Surveyor (scheduled 2008) could be able to determine the value of the energy density parameter with a precision of $1 \%$. A value lower than $1.009$ would discard the PDS as a model for cosmic space, in the sense that the size of the corresponding dodecahedron would become greater than the observable universe and would not leave any observable imprint on the CMB, whereas a value greater than $1.01$ would strengthen its cosmological pertinence.
\item If space has a non trivial topology, there must be particular correlations in the CMB, namely pairs of ``matched circles'' along which temperature fluctuations should be the same~\cite{CSS98}. The PDS model~\cite{Lum03} predicts 6 pairs of antipodal circles with an angular radius comprised between $5^{\circ}$ and $55^{\circ}$ (sensitively depending on the cosmological parameters).
\end{itemize} 

\begin{figure}[h!]
\centering
\includegraphics[width=11cm]{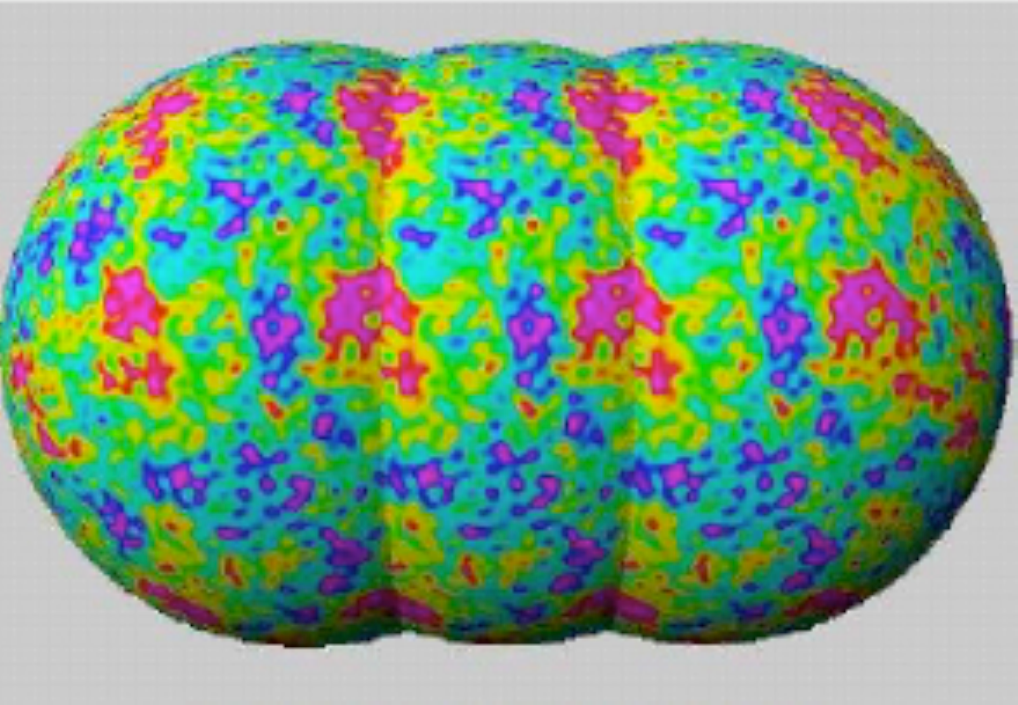}
\caption{ A multiply connected topology translates into the fact that any object in space may possess several copies of itself in the observable Universe. For an extended object like the region of emission of the CMB radiation we observe (the so-called last scattering surface) it can happen that it intersects with itself along pairs of circles. In this case, this is equivalent to say that an observer (located at the center of the last scattering surface) will see the same region of the Universe from different directions. As a consequence, the temperature fluctuations will match along the intersection of the last scattering surface with itself, as illustrated in the above figure. This CMB map is simulated for a multiconnected flat space - namely a cubic hypertorus whose length is 3.17 times smaller than the diameter of the last scattering surface. Only two duplicates are depicted.}
\label{figRia1}      
\end{figure}

\begin{figure}[h!]
\centering
\includegraphics[width=11cm]{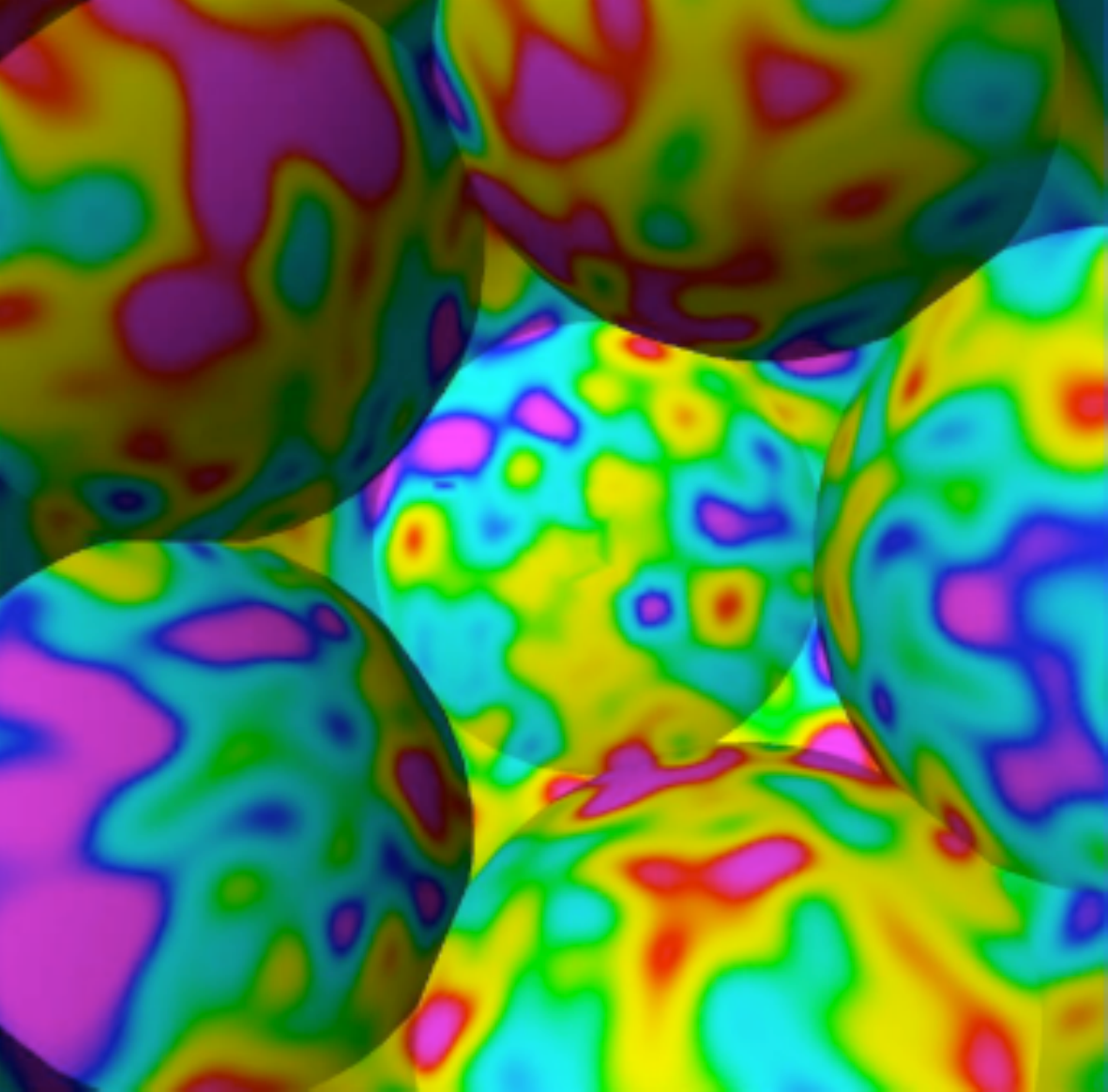}
\caption{ The last scattering surface seen from outside in the universal covering space of the Poincar\'e dodecahedral space with $\Omega_0 = 1.02$, $\Omega_m = 0.27$ and $h = 0.70$ (using modes up to a resolution of $6^{\circ}$). Since the volume of the physical space is about $80 \%$ of the volume of the last scattering surface, the latter intersects itself along six pairs of matching circles.}
\label{figRia2}      
\end{figure}

Such circles have been searched in WMAP data by several teams, using various statistical indicators and massive computer calculations. First, Cornish et al.~\cite{CSS04} claimed to have found no matched circles on angular sizes greater than $25^{\circ}$, and thus rejected the PDS hypothesis. Next, Roukema et al.~\cite{Roukema} performed the same analysis for smaller circles, and found six pairs of matched circles distributed in a dodecahedral pattern, each circle on an angular size about $11^{\circ}$. This implies $\Omega_0 = 1.010 \pm 0.001$ for $\Omega_m = 0.28 \pm 0.02$, values which are perfectly consistent with the PDS model. Finally, Aurich et al.~\cite{ALS1} performed a very careful search for matched circles and found that the putative topological signal in the WMAP data was considerably degraded by various effects~\cite{Then05}, so that the dodecahedral space model could be neither confirmed nor rejected... 

The controversy still went up a tone when Key et al.~\cite{CSS06} claimed that their negative analysis was not disputable, and that accordingly, not only the dodecahedral hypothesis was excluded, but also any multiply-connected topology on a scale smaller than the horizon radius. Since such an argument of authority, a fair portion of the academic community believes the WMAP data has ruled out multiply-connected universe models. However, at least the second part of the claim is wrong. The reason is that they searched only for antipodal or nearly-antipodal matched circles. But Riazuelo et al.~\cite{Ria04b} have shown that for generic multiply-connected topologies (including the well-proportioned ones, which are good candidates for explaining the WMAP power spectrum), the matched circles are generally not antipodal; moreover, the positions of the matched circles in the sky depend on the observer's position in the fundamental polyhedron. The corresponding larger number of degrees of freedom for the circles search in the WMAP data generates a dramatic increase of the computer time, up to values which are out-of-reach of the present facilities. It follows that the debate about the pertinence of a multiply connected well-proportioned space model to reproduce CMB observations is fully open. 

\begin{figure}[h!]
\centering
\includegraphics[height=6cm]{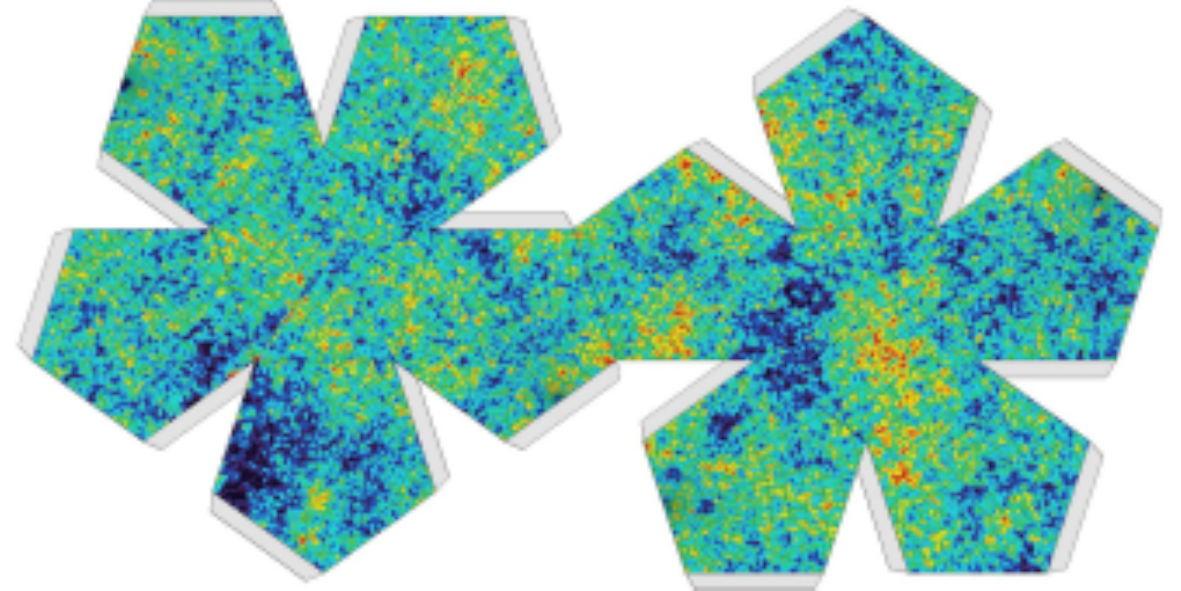}
\caption{ A pocket dodecahedral space, to be cut and glued by the reader (courtesy R. Lehoucq).}
\label{figRL}      
\end{figure}

The new release of WMAP data~\cite{WMAP3}, integrating two additional years of observation with reduced uncertainty, strengthened the evidence for an abnormally low quadrupole and other features which do not match with the infinite flat space model. Besides the quadrupole suppression, an anomalous alignment between the quadrupole and the octopole was put in evidence along a so-called ``axis of evil'' ~\cite{Lan05}. Thus the question arose to know whether, since non-trivial spatial topology can explain the weakness of the low-$\ell$ modes, might it also explain the quadrupole-octupole alignment? Until then no multiply-connected space model, either flat~\cite{Cres06} or spherical~\cite{ALST07},~\cite{WG07} was proved to exhibit the alignment observed in the CMB sky. This is not a strong argument against such models, since the  ``axis of evil'' is generally interpreted as due to local effects and foreground contaminations~\cite{Pru05}.

As a provisory conclusion, since some power spectrum anomalies are one of the possible signatures of a finite and multiply-connected universe, there is sill a continued interest in the Poincar\'e dodecahedral space and related finite universe models. And even if the particular dodecahedral space is eventually ruled out by future experiments, all of the other models of well-proportioned spaces will not be eliminated as such. In addition, even if the size of a multiply-connected space is larger (not too much) than that of the observable universe, we could all the same discover an imprint in the fossil radiation, even while no pair of circles, much less ghost galaxy images, would remain~\cite{Kunz07}. The topology of the universe could therefore provide information on what happens outside of the cosmological horizon! But this is search for the next decade \ldots

\printindex
\end{document}